\documentclass[preprint,showpacs,preprintnumbers,amsmath,amssymb,aps,showkeys]{revtex4}

\usepackage{graphicx}
\usepackage{dcolumn}
\usepackage{bm}
\usepackage{xspace}

\begin{document}


\title{Approximating the Largest Eigenvale of the Modified Adjacency Matrix of Networks with Heterogeneous Node Biases}

\author{Edward Ott}
 \email{edott@umd.edu}
\author{Andrew Pomerance}
\affiliation{%
Institute for Research in Electronics and Applied Physics \\
University of Maryland, College Park\\
College Park, MD, 20752
}%

\newcommand{\tp}[2]{\ensuremath{{#1}^{(#2)}} \xspace}
\newcommand{\lQ}[1][Q]{\ensuremath{\lambda_{#1}}\xspace}
\newcommand{\rQ}{\ensuremath{\rho_{Q}}\xspace}
\newcommand{\lQh}[1][Q]{\ensuremath{\hat{\lambda}_{#1}}\xspace}
\newcommand{\qi}[1][i]{\ensuremath{q_{#1}}\xspace}
\newcommand{\si}[1][i]{\ensuremath{\sigma_{#1}}\xspace}
\newcommand{\ex}[1]{\ensuremath{\langle #1 \rangle} \xspace}
\newcommand{\din}[1][]{\ensuremath{d^{in}_{#1}} \xspace}
\newcommand{\dout}[1][]{\ensuremath{d^{out}_{#1}} \xspace}
\newcommand{\douth}[1][]{\ensuremath{\hat{d}^{out}_{#1}} \xspace}
\newcommand{\dinh}[1][]{\ensuremath{\hat{d}^{in}_{#1}} \xspace}
\newcommand{\Zi}[1][i]{\ensuremath{\underline{z}_{#1}} \xspace}
\newcommand{\attr}[1][]{\ensuremath{(\Zi[#1], \si[#1])} \xspace}
\newcommand{\attrp}[1][]{\ensuremath{(\Zi[#1]', \si[#1]')} \xspace}
\newcommand{\attrpp}[1][]{\ensuremath{(\Zi[#1]'', \si[#1]'')} \xspace}
\newcommand{\ddist}{\ensuremath{P_{\si[]}(\Zi[])} \xspace}
\newcommand{\markov}[1][]{\ensuremath{\Pi(\Zi[]#1, \si[]#1 | \Zi[]#1', \si[]#1')} \xspace}
\newcommand{\markovsub}[2]{\ensuremath{\Pi(\Zi[#1], \si[#1] | \Zi[#2], \si[#2])} \xspace}
\newcommand{\markovZ}[1][]{\ensuremath{\Pi(\Zi[]#1 | \Zi[]#1')} \xspace}
\newcommand{\markovZa}[1][]{\ensuremath{\tp{\Pi}{0}(\Zi[]#1 | \Zi[]#1')} \xspace}
\newcommand{\markovZb}[1][]{\ensuremath{\tp{\Pi}{1}(\Zi[]#1 | \Zi[]#1')} \xspace}
\newcommand{\Ns}{\ensuremath{N_{\si[]}} \xspace}
\newcommand{\un}[1]{\tp{u}{#1}}
\newcommand{\vn}[1]{\tp{v}{#1}}
\newcommand{\pss}{\ensuremath{p_{\si[] \si[]'}} \xspace}
\newcommand{\Mss}{\ensuremath{M_{\si[] \si[]'}} \xspace}

\date{\today}
\begin{abstract}
Motivated by its relevance to various types of dynamical behavior of network systems, the maximum eigenvalue \lQ[A] of the adjacency matrix $A$ of a network has been considered, and mean-field-type approximations to \lQ[A] have been developed for different kinds of networks.  Here $A$ is defined by $A_{ij} = 1$ ($A_{ij} = 0$) if there is (is not) a directed network link to $i$ from $j$.  However, in at least two recent problems involving networks with heterogeneous node properties (percolation on a directed network and the stability of Boolean models of gene networks), an analogous but different eigenvalue problem arises, namely, that of finding the largest eigenvalue \lQ of the matrix $Q$, where $Q_{ij} = \qi A_{ij}$ and the `bias' \qi may be different at each node $i$.  (In the previously mentioned percolation and gene network contexts, \qi is a probability and so lies in the range $0 \le \qi \le 1$.)  The purposes of this paper are to extend the previous considerations of the maximum eigenvalue \lQ[A] of $A$ to \lQ, to develop suitable analytic approximations to \lQ, and to test these approximations with numerical experiments.  In particular, three issues considered are (i) the effect of the correlation (or anticorrelation) between the value of \qi and the number of links to and from node $i$; (ii) the effect of correlation between the properties of two nodes at either end of a network link (`assortativity'); and (iii) the effect of community structure allowing for a situation in which different $q$-values are associated with different communities.
\end{abstract}

\pacs{89.75.Hc}
\keywords{complex networks, eigenvalues and eigenfunctions, matrix algebra, genetic networks, Boolean networks, percolation}
\maketitle

\section{\label{sec:introduction}Introduction}

Topological properties of networks have received much attention.  The coarsest topological property is the degree distribution, where the in-degree \din[i] and out-degree \dout[i] of a network node $i$ are defined as the number of directed network links pointing into and away from node $i$.  At each node, the in-degree and out-degree may be correlated; this correlation can be characterized by 
\begin{equation}
\eta = \ex{\din\dout}/\ex{d}^2,
\end{equation}
where \ex{\cdot} denotes the average of the indicated quantity over all network nodes.  (Since every out-link of a node is an in-link for some other node, $\ex{\din} = \ex{\dout} \equiv \ex{d}$.)   Additionally, networks can be assortative or disassortative \cite{bib1}, i.e., nodes with high degree may prefer or avoid connecting to other nodes of high degree.  We characterize this by a correlation coefficient $\rho$ between the in-degrees \din[j] and the out-degrees \dout[i] at either end of a directed link from node $j$ to node $i$ \cite{bib2, bib3},
\begin{equation}
\label{eq:2}
\rho = \ex{\din[i]\dout[j]}_e/(\eta\ex{d})^2,
\end{equation}
where $\ex{\cdot}_e$ indicates an average over all network links.  If $\rho > 1$, the network is assortative, if $\rho < 1$ it is disassortative, and it is neutral if $\rho$ is exactly one.  Another example of topological structure is the existence of communities \cite{bib4}, which are groups of nodes that tend to be densely interconnected within the group but sparsely connected between groups.

In addition to static topological properties of networks, another area of recent interest has been dynamical processes taking place on networks.  Examples include synchronization of coupled identical dynamical systems (which may be chaotic) \cite{bib5}, the onset of coherence in the evolution of heterogeneous network-coupled dynamical systems (both oscillatory \cite{bib6}, as well as chaotic \cite{bib7}), the onset of instability in discrete state models of gene networks \cite{bib8}, percolation on directed networks \cite{bib9}, and others.  In several of these examples \cite{bib7, bib8, bib9}, an important determining quantity was shown to be the maximum eigenvalue of the adjacency matrix $A$, the elements of which are defined to be $A_{ij} = 1$ if there is a directed link to node $i$ from node $j$ and zero otherwise for all $i$, $j = 1, 2, ...,N$, where $N$ is the number of network nodes; $A_{ii} \equiv 0$ by definition.

Motivated by the importance of \lQ[A] to these dynamical problems, Ref. \cite{bib2} developed theory for obtaining large $N$ approximations to \lQ[A] from knowledge of statistical characterizations of the network.  For example, it was shown \cite{bib2} that random networks constrained only by specification of the joint degree distribution $P(\din, \dout)$ have 
\begin{equation}
\label{eq:lAcorr}
\lQ[A] \approx \ex{\din \dout}/\ex{d} = \eta \ex{d},
\end{equation}
where $P(\din, \dout)$ is the probability that a randomly chosen node has in-degree \din and out-degree \dout.  From \eqref{eq:lAcorr} it is seen that nodal correlation or anticorrelation between \din and \dout increases or decreases \lQ[A].  If an additional assortativity constraint is imposed, then Ref. \cite{bib2} obtains 
\begin{equation}
\label{eq:lAass}
\lQ[A] \approx \rho \eta \ex{d},
\end{equation}
to lowest order in $(\rho-1)$.  Thus assortativity ($\rho > 1$) tends to increases \lQ[A], and disassortativity ($\rho < 1$) tends to decrease \lQ[A].

However, we note that in two of the cited applications (namely, gene network stability \cite{bib8} and site percolation on large directed networks \cite{bib9}), the formulations resulted in a somewhat more general eigenvalue problem.  Specifically, this problem was that of determining the largest eigenvalue \lQ of a generalized adjacency matrix $Q$ whose elements are given by
\begin{equation}
\label{eq:Qdef}
Q_{ij} = q_i A_{ij},
\end{equation}
where \qi is the `bias' characterizing node $i$ which may be different for each node.  In the special case of uniform $\qi \equiv q$ for all $i$, the problem for \lQ reduces to that for \lQ[A], i.e., $\lQ = q\lQ[A]$, and the previous results such as Eqs. \eqref{eq:lAcorr} and \eqref{eq:lAass} can be employed.  However, it is also of interest to consider the more general problem of determining \lQ for nonuniform biases, and it is that problem to which this paper is devoted.  We note that in the site percolation context \cite{bib9}, $q_i = (1-p_i)$, where $p_i$ is the probability that node $i$ is removed, while in the gene network context \cite{bib8}, $q_i$ is the probability that the output of gene (node) $i$ is switched to another state if one or more of its randomly chosen inputs is switched.  In either case, $q_i$ is in the range $0 \leq \qi \leq 1$, and we accordingly restrict ourselves to $q_i \geq 0$.  Since $Q_{ij} \geq 0$, the Frobenius-Perron theorem implies that \lQ is real and positive.

Our analysis will consider `Markovian' random networks (see Sec. \ref{sec:markov}).  This type of consideration was used in the analysis of \lQ[A] in Ref. \cite{bib2}, as well as in a variety of other interesting studies of different network related problems (e.g., Refs. \cite{bib11, bib12, bib13, bib14} which consider epidemic spreading and percolation).  Basically, a Markovian network is one for which the only nontrivial spatial correlations are between nodes that are directly connected by a single link.  Within this framework, we formulate a theory for determining the \lQ of large networks, and we utilize our theory to examine several significant situations of interest.  Examples of our results are a generalization of Eq. \eqref{eq:lAcorr} (see Eq. \eqref{eq:Qeigcorr}) showing that correlation between $q$ and $\din \dout$ increases \lQ, a generalization of Eq. \eqref{eq:lAass} (see Eq. \eqref{eq:eigass}) showing that correlation between $q_iq_j$ and $\dout[i]\din[j]$ on edges from $j \rightarrow i$ increases \lQ, and an analysis of the effect on \lQ of network communities tending to have different, community-dependent $q$ values (see Eq. \ref{eq:eigcomm}).

For later reference we note the following relationships involving the adjacency matrix,
\begin{eqnarray}
\label{eq:6}
\din[j] &=& \sum_{i=1}^{N} A_{ij},\; \dout[i] = \sum_{j=1}^{N}A_{ij}, \\
\label{eq:7}
\ex{S_{ij}}_e &=& \left[\sum_{i,j}A_{ij}S_{ij}\right]/\sum_{i,j} A_{ij}.
\end{eqnarray}
By \eqref{eq:6}
\begin{equation}
\label{eq:8}
\ex{\din} = \ex{\dout} = \frac{1}{N}\sum_{i,j}A_{ij} \equiv \ex{d}.
\end{equation}
By \eqref{eq:6}-\eqref{eq:8}
\begin{eqnarray}
\label{eq:9}
\nonumber \ex{\dout[i]}_e &=& \left[ \sum_{i,j} A_{ij}\dout[i] \right] / \sum_{i,j} A_{ij} = \sum_i \dout[i] \din[i] / \sum_{i,j} A_{ij} \\
 &=& \ex{\din\dout}/\ex{d}.
\end{eqnarray}

\section{\label{sec:markov}Markovian Networks}

We characterize each node $i$ by four attributes: its in-degree \din[i], its out-degree \dout[i], its `bias' \qi, and its group (or community), labeled \si.  We call the triplet $\Zi = (\din[i], \dout[i], \qi)$ the `generalized degree' of node $i$.  The number of groups is denoted $s$, so that $\si[] = 1, 2, ..., s$.  If no two nodes have the same attributes $\attr$, then there is a one-to-one correspondence between $i$ and $\attr \in \left\{ \attr[k] | k = 1,2,...N \right\}$.  We consider $N$-node random networks specified by the following quantities: (i) the number of nodes in each group $N_\sigma$ ($\sum_{\sigma = 1}^s N_\sigma = N$); (ii) the degree distribution \ddist for group \si[] ($\si[] = 1, 2, ..., s$) giving the probability that a node randomly chosen from group \si[] has generalized degree \Zi[]; and (iii) the probability \markov that, if a randomly chosen link originates from a node in group $\si[]'$ that has degree $\Zi[]'$, then that link points to a node in group \si[] with degree \Zi[].  Note that, since every out-link for a node is an in-link for some other node, the degree distributions \ddist are constrained to satisfy the relation,
\begin{equation}
\sum_{\Zi[], \si[]} \Ns \ddist \dout = \sum_{\Zi[], \si[]} \Ns \ddist \din,
\end{equation}
which we denote $N \ex{d}$.  Furthermore, we have that \markov satisfies the probability normalization condition,
\begin{equation}
\label{eq:markovnorm}
\sum_{\Zi[], \si[]} \markov = 1.
\end{equation}

By use of this model, we essentially assume that the only non-trivial correlation between the attributes of two different nodes occurs when they are directly connected by a single link.  For example, if we choose a random outward path of length two from a node in group \si[a] of degree \Zi[a], then the probability that the first leg of the path goes to a node having \attr[b], and the second leg of the path goes to a node having \attr[c] is given by $\markovsub{c}{b}\markovsub{b}{a}$.

In order to find the maximum eigenvalue of $Q$, we consider the iteration $\un{n+1} = Q\un{n}$ which, for a typical initial choice of \un{0} converges on the eigenvector $u$ corresponding to the largest eigenvalue \lQ.  Relabeling the nodes by their attributes \attr, we write the components of the vector $u = \left[ u_1, u_2, ..., u_N \right]^T$ as $u_i = v\attr[i]$.  The ensemble average of the iterated vector $\vn{n}\attr$ thus evolves according to
\begin{equation}
\vn{n}\attr = q \sum_{\si[]'} \sum_{\Zi[]'} \markov (\dout)' \vn{n}\attrp,
\end{equation}
and we denote the eigenvalue of this evolution by $\lQh$,
\begin{equation}
\label{eq:eighat}
\lQh v\attr = q \sum_{\si[]'} \sum_{\Zi[]'} \markov (\dout)' v\attrp.
\end{equation}

For large $N$, and a random draw from our Markov ensemble of networks, we suppose that $\lQh$ from \eqref{eq:eighat} will typically provide a good approximation to \lQ for the chosen network, and we will test this supposition using numerical experiments.  In the next section, we apply Eq. \eqref{eq:eighat} to obtain analytical approximations to \lQ for several situations of interest.

\section{Evaluation of \lQ\label{sec:lQ}}
\subsection{The Effect of Nodal Correlations\label{sec:correlation}}
We first consider the case $s = 1$, corresponding to the absence of group structure.  Thus the variable \si[] may be omitted from Eq. \eqref{eq:eighat}.  We furthermore assume that \Zi and \Zi[j] on the two ends of a link from $j$ to $i$ are uncorrelated.  Thus there is no assortativity, and \markovZ does not depend on $\Zi[]'$.  Under this assumption, \markovZ is simply the probability that a randomly chosen link points toward a node with degree \Zi[].  This probability is proportional to the number of nodes with degree \Zi[], and to the number of in-links to such a node, 
\begin{equation}
\label{eq:markovZ}
\markovZ = \din P(\Zi[]) / \ex{d},
\end{equation}
where the factor $\ex{d}^{-1}$ provides the necessary normalization from Eq. \eqref{eq:markovnorm}.  Inserting \eqref{eq:markovZ} into \eqref{eq:eighat} we have that 
\begin{equation}
\label{eq:correig}
\lQh v(\Zi[]) = q \din P(\Zi[])\ex{d}^{-1} \sum_{\Zi[]'}(\dout)'v(\Zi[]').
\end{equation}
Thus we see that the eigenvector $v(\Zi[])$ is 
\begin{equation}
\label{eq:veccorr}
v(\Zi[]) = q \din P(\Zi[]),
\end{equation}
which when inserted into \eqref{eq:correig} yields 
\begin{equation}
\label{eq:Qeigcorr}
\lQh = \ex{q \din \dout} / \ex{d},
\end{equation}
where 
\begin{equation}
\ex{q \din \dout} = \sum_{\Zi[]} q \din \dout P(\Zi[]).
\end{equation}
Equation \eqref{eq:Qeigcorr} is the appropriate generalization of Eq. \eqref{eq:lAcorr} to take into account the node-dependent biases \qi that appear in the definition, Eq. \eqref{eq:Qdef}, of the matrix $Q$.

If $q$ and $\din \dout$ are uncorrelated, $\lQh = \ex{q} \lQh[A]$ where \lQh[A] is given by \eqref{eq:lAcorr}.  On the other hand, we see that if $q$ and $\din \dout$ are correlated (anticorrelated), then \lQh is larger (smaller) than $\ex{q} \lQh[A]$.

\subsection{Assortativity\label{sec:assortativity}}

Next we wish to consider how the result in Eq. \eqref{eq:Qeigcorr} is modified if we allow correlation between \Zi[] and $\Zi[]'$.  We address this problem perturbatively, and we write \markovZ as 
\begin{equation}
\label{eq:markovexp}
\markovZ \approx \markovZa + \epsilon \markovZb,
\end{equation}
where $\epsilon$ is a small expansion parameter, and \markovZa is given by the uncorrelated result, Eq. \eqref{eq:markovZ}.  Similarly expanding the eigenvalue \lQh and the eigenvector $v(\Zi[])$, we have 
\begin{eqnarray}
\label{eq:eigexp}
\lQh &\approx& \tp{\lQh}{0} + \epsilon \tp{\lQh}{1}, \\
\label{eq:vecexp}
v(\Zi[]) &\approx& \vn{0}(\Zi[]) + \epsilon \vn{1}(\Zi[]),
\end{eqnarray}
where \tp{\lQh}{0} is given by \eqref{eq:Qeigcorr} and $\vn{0}(\Zi[])$ is given by \eqref{eq:veccorr}.  Inserting \eqref{eq:markovexp}-\eqref{eq:vecexp}, \eqref{eq:markovZ}, \eqref{eq:veccorr}, and \eqref{eq:Qeigcorr} into \eqref{eq:eighat}, multiplying the resulting equation by \dout, and summing over all \Zi[], the terms involving $\vn{1}(\Zi[])$ cancel.  Thus we obtain
\begin{equation}
\label{eq:21}
\epsilon \tp{\lQh}{1} \tp{\lQh}{0} = \epsilon \sum_{\Zi[], \Zi[]'} q \dout (\din)' q' \markovZb \tilde{P}(\Zi[]),
\end{equation}
where
\begin{equation}
\label{eq:22}
\tilde{P}(\Zi[]) = \dout P(\Zi[])/\ex{d}.
\end{equation}
is the probability that a randomly chosen link originates from a node of generalized degree \Zi[].  With this interpretation of \eqref{eq:22}, we see that \eqref{eq:21} can be re-expressed in terms of the link average $\ex{\cdot}_e$,
\begin{equation}
\label{eq:23}
\epsilon \tp{\lQh}{1} \tp{\lQh}{0} = \epsilon \ex{\qi \dout[i] \din[j] \qi[j]}_e - \ex{\qi \dout[i]}_e \ex{\qi[j]\din[j]}_e,
\end{equation}
where we use the convention that $j$ ($i$) labels the node that the link comes from (points to).

Proceeding as in Eq. \eqref{eq:9}, we obtain
\begin{equation}
\label{eq:24}
\ex{\qi \din[i]}_e = \ex{\qi[j] \din[j]}_e = \ex{q \din \dout}/\ex{d} = \tp{\lQh}{0},
\end{equation}
which when inserted in \eqref{eq:23} yields
\begin{equation}
\label{eq:25}
\lQh \approx \tp{\lQh}{0} + \epsilon \tp{\lQh}{1} = \ex{\qi \dout[i] \din[j] \qi[j]}_e / \tp{\lQh}{0}.
\end{equation}
Now defining a new assortativity coefficient appropriate to networks with heterogeneous biases \qi, we write
\begin{equation}
\label{eq:rhoQ}
\rQ = \frac{\ex{\qi \dout[i] \din[j] \qi[j]}_e}{\ex{\qi \dout[i]}_e \ex{\qi[j]\din[j]}_e} = \frac{\ex{\qi \dout[i] \din[j] \qi[j]}_e}{(\tp{\lQh}{0})^2},
\end{equation}
in terms of which Eq. \eqref{eq:25} takes the suggestive form,
\begin{equation}
\label{eq:eigass}
\lQh \approx \tp{\lQh}{0} \rQ.
\end{equation}
Thus bias assortativity (disassortativity), corresponding to $\rQ > 1$ ($\rQ < 1$) yields $\lQh > \tp{\lQh}{0}$ ($\lQh < \tp{\lQh}{0}$).  Equations \eqref{eq:rhoQ} and \eqref{eq:eigass} generalize Eqs. \eqref{eq:2} and \eqref{eq:lAass} for \lQ[A] to results for \lQ.

\subsection{Community and Bipartite Structure\label{sec:community}}

We now consider how the presence of several network groups ($s > 1$) influence \lQh.  As in Sec. \ref{sec:correlation}, we assume that \Zi is uncorrelated with \Zi[j], where \Zi and \Zi[j] are at either end of a link from $i$ to $j$.  However, we do include correlations between \si and \si[j] along this link, and we characterize this correlation by the $s \times s$ matrix of transition probabilities \pss, giving the probability that a randomly chosen out-link from a node in group $\si[]'$ connects to a node in group \si[].  With these assumptions, we have the following result for \markov (analogous to \eqref{eq:markovZ}), 
\begin{equation}
\label{eq:28}
\markov = D^{-1}(\si[]') \pss \din \ddist,
\end{equation}
where $D(\si[]') = \sum_{\Zi[], \si[]} \ddist \din \pss$ is a normalizing factor (see Eq. \eqref{eq:markovnorm}).  Inserting \eqref{eq:28} into \eqref{eq:eighat}, 
\begin{equation}
\label{eq:29}
\lQh v\attr = q \sum_{\si[]', \Zi[]'} D^{-1}(\si[]') \din \pss \ddist (\dout)' v\attrp.
\end{equation}
Equation \eqref{eq:29} immediately determines the \Zi[] dependence of $v\attr$.  Thus we can write
\begin{equation}
\label{eq:30}
v\attr = q\din \ddist w(\si[]),
\end{equation}
where the \si[] dependent quantity $w(\si[])$ is, as yet, undetermined.  Substituting \eqref{eq:30} into \eqref{eq:29} we obtain the following eigenvalue equation for $w(\si[])$ and \lQh,
\begin{equation}
\label{eq:31}
\lQh w(\si[]) = \sum_{\si[]} \Mss w(\si[]'),
\end{equation}
where $M$ is the $s \times s$ matrix,
\begin{equation}
\label{eq:32}
\Mss = D^{-1}(\si[]') \ex{q \din \dout}_{\si[]'} \pss,
\end{equation}
where $\ex{\cdot}_{\si[]} = \sum_{\Zi[]} ( \cdot ) \ddist$.  Thus the $N \times N$ eigenvalue problem for \lQ is now approximated by the much smaller $s \times s$ eigenvalue problem \eqref{eq:31},
\begin{equation}
\label{eq:eigcomm}
\lQh = \text{max. eigenvalue} [M].
\end{equation}
We have also expanded the group eigenvalue problem to obtain the correction to \eqref{eq:eigcomm} that is introduced by including correlations between \Zi and \Zi[j] along links from $j$ to $i$.  This analysis proceeds in a manner similar to that in Sec. \ref{sec:assortativity} and is omitted.

Note that in the case where the off-diagonal transition probabilities are zero, $\pss = 0$ for $\si[] \neq \si[]'$, we have $s$ completely disconnected groups, and that, for $\pss/N_{\si[]}$ independent of \si[] and $\si[]'$, the group-dependence on the connectivity is absent.  If the diagonal terms of the matrix $\pss/N_{\si[]}$ are larger than the off-diagonal terms, then we say there is `community structure' (i.e., the density of intragroup connections is larger than the density of intergroup connections).  

At the opposite extreme, for the case of two groups ($s = 2$), if the diagonal components of the transition probability matrix are zero ($p_{\si[]\si[]} = 0$ for $\si[] = 1,2$), then connections exist only between, and not within, the two groups, i.e., the network is `bipartite.'  Thus, if the diagonal terms of the matrix  $\pss/N_{\si[]}$ are smaller than the off-diagonal terms, then we say the network has `bipartite structure.'

In our numerical tests of Eq. \eqref{eq:eigcomm} in Sec. \eqref{sec:community_test}, we will consider two groups ($s = 2; \si[] = 1,2$) with equal sizes ($N_1 = N_2 = N/2$) and with symmetric transition properties ($p_{11} = p_{22} \equiv p_0$, $p_{12} = p_{21} \equiv p_x$).  We will, in addition, restrict our consideration to the case where the in-degree/out-degree distributions are the same for the two groups, but we will allow the biases $q$ for the two groups to be unequal, with the $q$'s not correlated with \din and \dout; i.e.,
\begin{equation}
\label{eq:ddistcomm}
\ddist = P^d(\din, \dout) P_{\si[]}^q(q).
\end{equation}
With these conditions Eq. \eqref{eq:eigcomm} reduces to 
\begin{equation}
\Mss = \ex{q}_{\si[]'}\xi\pss,
\end{equation}
where
\begin{equation}
\xi = D^{-1}(1)\ex{\din \dout}_1 = D^{-1}(2)\ex{\din \dout}_2,
\end{equation}
or
\begin{equation}
\label{eq:M}
M = \xi \left[ \begin{array}{cc}
p_0\ex{q}_1 & p_x\ex{q}_1 \\
p_x\ex{q}_2 & p_0\ex{q}_2
\end{array} \right].
\end{equation}

From Eqs. \eqref{eq:eigcomm} and \eqref{eq:M},
\begin{equation}
\label{eq:38}
\lQh = \xi\left\{ p_0(\ex{q}_1 + \ex{q}_2) + \left[ p_0(\ex{q}_1 - \ex{q}_2)^2 + 4p_x\ex{q}_1\ex{q}_2)\right]^{1/2} \right\}/2.
\end{equation}
Equation \eqref{eq:38} can be put in a somewhat more revealing form by introducing $q_{\pm} = (\ex{q}_1 \pm \ex{q}_2)/2$, in terms of which \eqref{eq:38} becomes
\begin{equation}
\label{eq:eigcomm2}
\lQh = \xi\left\{p_0 q_+ + \left[(p_0^2-p_x^2)q_-^2 + p_x^2 q_+^2 \right]^{1/2}\right\}.
\end{equation}
From \eqref{eq:eigcomm2} we see that, if we keep $q_+$ (the average $q$ value for the whole network) fixed, but allow the difference between the average $q$'s in the two groups to increase (i.e., we increase $|q_-|$), then \lQh increases if the network has community structure ($p_0 > p_x$), but it decreases if the network has bipartite structure ($p_x > p_0$).

\section{Numerical Tests\label{sec:tests}}

\subsection{First-order Approximation}
We test the predictions of Eq. \eqref{eq:Qeigcorr} on networks with equal power-law in-degree and out-degree distributions.  To construct the networks used to test this hypothesis, we follow the method used in \cite{bib2}.  In particular, we first randomly construct a list of $N$ degree values by choosing $N$ random numbers drawn from a given distribution, in this case,
\begin{equation}
\label{eq:degreedist}
P(d) \propto \begin{cases}
d^{-\gamma}, 						& d^{min} \leq d \leq d^{max},\\
0, 	& \text{otherwise}.
\end{cases}
\end{equation}
We use $\gamma = 2.5$ and adjust $d^{min}$ and $d^{max}$ to tune \ex{d}.  We then assign each number on this list to each node $i$ and we call this assignment the `target' in-degree \dinh[i].  Next we use this same list to assign to each node $i$ a target out-degree \douth[i], and perform this assignment in one of three ways: (i) $\douth[i] = \dinh[i]$, yielding maximal $\ex{\din \dout}$ and $\eta$; (ii) \douth[i] randomly drawn from the list, yielding $\ex{\din \dout} \approx \ex{d}^2$ and $\eta \approx 1$; or (iii) the node with the largest \dinh[] is assigned \douth equal to the smallest value on the list, the node with the second largest \dinh is assigned \douth equal to the second smallest value on the list, etc., yielding minimal $\ex{\din \dout}$ and $\eta$.  Once the \dinh and \douth are assigned, we construct the network by setting the elements of the adjacency matrix $A_{ij} = 1$ with probability $\dinh[i]\douth[j]/N\ex{\hat{d}}$ (where $\ex{\hat{d}} = \ex{d}$ is the average of the list values) and 0 otherwise.

\begin{figure}
\includegraphics{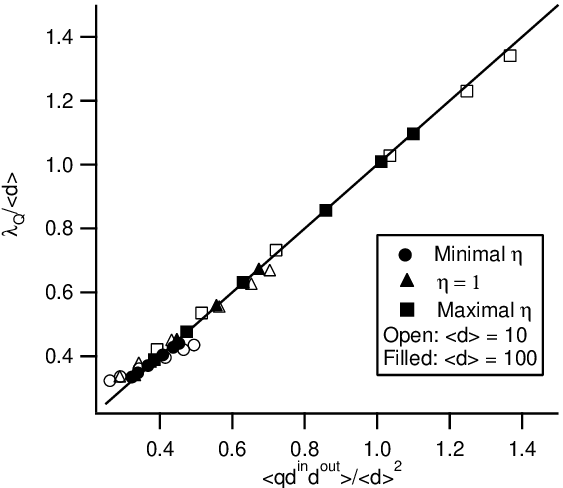}
\caption{\lQ/\ex{d} vs. $\ex{q \din\dout}/\ex{d}^2$ for networks of size $N=10^4$ with no assortativity and $\ex{d} = 100$ (filled markers) and $\ex{d} = 10$ (open markers).  For both values of $\ex{d}$, three values of $\eta$ are considered: maximal (circle), neutral (triangle) and minimal (squares).  Each marker is the average of 10 networks, and the solid line is the theoretical prediction, $\lQ/\ex{d} = \ex{q \din\dout}/\ex{d}^2$.}
\label{fig:corr}
\end{figure}

After the network is constructed, we assign the biases \qi drawn from a uniform distribution on the interval $[0,1].$  For each of the three values of $\ex{ \din \dout }$, we tune $\ex{q \din \dout}$ by swapping the biases of random pairs of nodes to increase or decrease $\ex{q \din \dout}$.  For example, if we wish to obtain increased \ex{q \din \dout},  we only keep those swaps that increase $\ex{q \din \dout}$.  In Fig. \ref{fig:corr}, we plot the measured normalized eigenvalue $\lQ/\ex{d}$ vs. $\ex{q \din\dout}/\ex{d}^2$ for $\ex{d} \approx 10$ (open markers) and $\ex{d} \approx 100$ (filled markers) for $\ex{\din \dout}$ maximal (circles), minimal (squares), and neutral (triangles) averaged over 10 networks.  As can be seen, the markers all fall on the solid line, $\lQ/\ex{d} = \ex{q \din\dout}/\ex{d}^2$.

\subsection{Assortativity}

\begin{figure}
\includegraphics{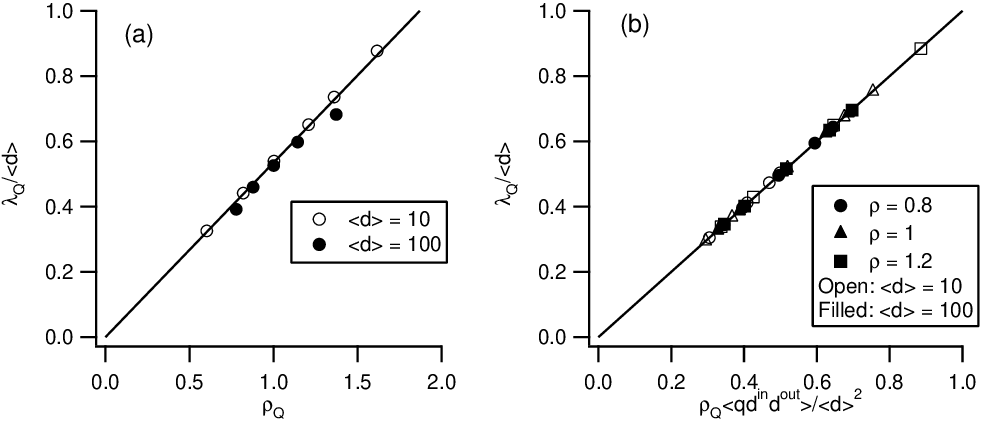}
\caption{(a) $\lQ/\ex{d}$ vs. \rQ for a network of size $N = 10000$ and average connectivity $\ex{d} = 10$ (circles) and $\ex{d} = 100$ (squares).  Each marker is the average of 10 networks. (b)$\lQ/\ex{d}$ vs. $\tp{\lQh}{0}\rQ/\ex{d}$ for networks with structural assortativity of 0.8 (circles), 1.0 (triangles), and 1.2 (squares) of size $N = 10000$ and average connectivity $\ex{d} = 10$ (open markers) and $\ex{d} = 100$ (filled markers).}
\label{fig:ass}
\end{figure}

In Fig. \ref{fig:ass} we test the predictions of Eq. \eqref{eq:eigass}.  Baseline networks of size $N = 10^4$ with $\eta \approx 1$ and $\rho \approx 1$ are constructed as described above, with the biases \qi drawn from a uniform distribution on the interval $[0,1]$.  We then consider two methods of tuning \rQ, method (a), which yields networks with $\rQ \neq 1$ but no correlation between the degrees and the bias at a given node, and method (b), which introduces nodal degree-bias correlations.  

\textit{Method (a).}  This method is a modified version of the algorithm in \cite{bib2}: (i) Randomly choose two links going from $j_1 \rightarrow i_1$ and $j_2 \rightarrow i_2$.  (ii) Calculate $d^{in}_{j_1}d^{out}_{i_1}\qi[j_1]\qi[i_1]+d^{in}_{j_2}d^{out}_{i_2}\qi[j_2]\qi[i_2]$ and $d^{in}_{j_1}d^{out}_{i_2}\qi[j_1]\qi[i_2]+d^{in}_{j_2}d^{out}_{i_1}\qi[j_2]\qi[i_1]$.  (iii)  If the latter value is larger or smaller (depending on whether the target \rQ is greater or less than one), delete the original links and place new links from $j_1 \rightarrow i_2$ and $j_2 \rightarrow i_1$, otherwise keep the original links.  (iv)  Repeat this process until the target \rQ is achieved.  

\textit{Method (b).}  This method is a two-step process.  First, we tune $\rho$ by swapping inputs of random link pairs, and we do this without regard to the node biases, as in \cite{bib2}, yielding a network with $\rQ \approx \rho$.  Once $\rho$ is tuned, we futher tune the bias assortativity \rQ by the following: (i) Randomly choose two nodes, $i$ and $j$.  (ii)  Calculate the change in \rQ that would result if the $q$'s at these two randomly chosen nodes were interchanged.  (iii)  If it is desired to increase (decrease) \rQ and the change in \rQ is positive (negative), then swap the $q$ values; otherwise do not make the swap.  (iv) Repeat the above process until the target \rQ is achieved.

The results of these two methods are in Fig. \ref{fig:ass}.  Each marker in the figure is the average of 10 networks of size $N = 10^4$, and we consider networks with $\ex{d} = 10$ (open markers) and $\ex{d} = 100$ (filled markers) tuned with both methods.  In Fig. \ref{fig:ass}(a), we plot the normalized eigenvalue $\lQ/\ex{d}$  vs. \rQ for networks tuned with method (a).  Since $\eta$ is approximately unity and the \qi are assigned independently of the node degrees, $\tp{\lQh}{0} \approx \ex{d}\ex{q}$; the theoretical prediction (solid curve) is therefore $\lQh/\ex{d} = \rQ\ex{q}$.  The results of method (b) are shown in Fig. \ref{fig:ass}(b).  We consider networks tuned to $\rho \approx 0.8$ (circles), 1.0 (triangles), and 1.2 (squares).  Note that swapping the \qi[] values in method (b) changes \tp{\lQh}{0} as well as \rQ, and thus \tp{\lQh}{0} must be calculated for every marker.  We therefore plot \lQ/\ex{d} vs. $\tp{\lQh}{0}\rQ/\ex{d}$, and we see that all the points fall on the theoretical prediction, $\lQ/\ex{d} = \tp{\lQh}{0}\rQ$/\ex{d}.

\subsection{Community Structure\label{sec:community_test}}

\begin{figure}
\includegraphics{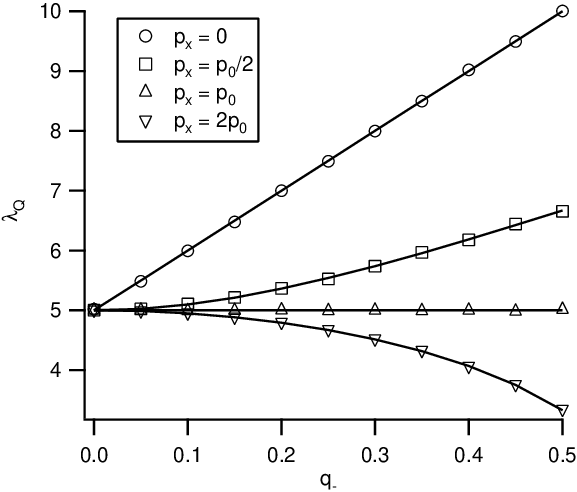}
\caption{\lQ vs. $q_-$ for networks of size $N = 10^4$ with two equal groups of varying type: two completely separated components ($p_x = 0$, circles), strong community structure ($p_x = p_0/2$, squares), no group structure ($p_x = p_0$, upward pointing triangle), and strong bipartite structure ($p_x = 2p_0$, downward pointing triangle).  Each marker is the average of 10 networks, and the solid curves are the theoretical predictions of Eq. \eqref{eq:eigcomm2}.}
\label{fig:comm}
\end{figure}

In Fig. \ref{fig:comm} we test the predictions of the community structure theory for a network of size $N = 10^4$ with two equally sized groups.  Networks are constructed as described Sec. \ref{sec:community}; $A_{ij} = 1$ with probability $p_0$ if $j$ and $i$ are in the same group or probability $p_x$ if $i$ and $j$ are in different groups.  We then consider four cases: two completely separated components ($p_x = 0$, circles), strong community structure ($p_x = p_0/2$, squares), no group structure ($p_x = p_0$, upward pointing triangles), and strong bipartite structure ($p_x = 2p_0$, downward pointing triangles).  The groups have uniform biases, $q_1 = q_+ + q_-$ and $q_2 = q_+ - q_-$, with $q_+ = 0.5$ and $q_-$ varying from 0 to 0.5.  We plot the measured \lQ vs. the difference in group biases, $q_-$, averaged over 10 networks.  The solid curves are the theoretical predictions of Eq. \eqref{eq:eigcomm2}, and markers are the average of 10 networks.

Again we obtain excellent agreement between the theory and the numerical tests.  Note that, as mentioned in Sec. \ref{sec:community}, the effect of increasing $q_-$ is to increase \lQ in the case with community structure and to decrease \lQ in the case with bipartite structure.

\section{Conclusion}

Motivated by recent work on the stability of gene network models \cite{bib8} and on percolation on directed networks \cite{bib9}, we have developed and numerically tested theoretical predictions for the maximum eigenvalues \lQ of the modified adjacency matrix $Q$ defined by Eq. \eqref{eq:Qdef}.  Using a Markov network model (Sec. \ref{sec:markov}), we calculate approximations to \lQ for various situations (Sec. \ref{sec:lQ}).  In particular, we considered: (i) the effect of correlation between the bias $q$ at a node with the product $\din[]\dout[]$ at that node; (ii) the effect of correlations between the degrees and biases for nodes at the two ends of a network link; and (iii) the effect of the existance of groups of nodes with community or bipartite structure in which different node bias distributions apply to different groups.  We find that the effects discussed strongly influence the value of \lQ, and in all cases our numerical tests (Sec. \ref{sec:tests}) resulted in excellent agreement with our theoretical results.

This work was supported by NSF (Physics) and by ONR (contract N00014-07-1-0734).  The work of A.P. was partly supported by the NCI intramural program.


\begin{thebibliography}{10}
\bibitem{bib3}
This definition of $\rho$ appears in Ref. \cite{bib2}.  Newman \cite{bib1} characterizes assortativity using a different quantity, but Eq. \eqref{eq:2} is more relevant for our purposes.
\end{thebibliography}
\end{document}